\documentclass[aps,prl,twocolumn,floatfix,showpacs,superscriptaddress]{revtex4}
\usepackage{epsfig}
\usepackage{amsmath}
\usepackage{hyperref}
\begin{document}
\def\il{I_{low}}
\def\iu{I_{up}}
\def\eeq{\end{equation}}
\def\ie{i.e.}
\def\etal{{\it et al. }}
\def\prb{Phys. Rev. {B }}
\def\pra{Phys. Rev. {A }}
\def\prl{Phys. Rev. Lett. }
\def\pla{Phys. Lett. A }
\def\ssc{Solid State Commun.}
\def\ajp{Am. J. Phys. }
\def\mpl{Mod. Phys. Lett. {B }}
\def\ijmpb{Int. J. Mod. Phys. {B }}
\def\ijp{Ind. J. Phys. }
\def\ijpap{Ind. J. Pure Appl. Phys. }
\def\ibm{IBM J. Res. Dev. }
\def\pjp{Pramana J. Phys.}
\def\pt{Phys. Today}
\def\epl{Euro. Phys. Lett.}
\def\jpcm{J. Phys. Condensed Matter }
\def\epjb{Eur. Phys. J. B}

\title{Crossed Andreev reflection as a probe for the pairing symmetry of Ferromagnetic Superconductors}

\author{Colin Benjamin}
\affiliation{C/M-24, VSS Nagar, Bhubaneswar-751007, Orissa,
India.}\altaffiliation{Present Address: Centre de Physique
Theorique, CNRS.UMR 6207-Case 907, Faculte
 des Sciences de Luminy, 13288 Marseille Cedex 09, France} \email{cbiop@yahoo.com}

\begin{abstract}
The coexistence of superconductivity and ferromagnetism has
brought about the phenomena of ferromagnetic superconductors. The
theory needed to understand the compatibility of such antagonistic
phenomena cannot be built until the pairing symmetry of such
superconductors is correctly identified. The proper and
unambiguous identification of the pairing symmetry of such
superconductors is the subject of this paper. This work shows that
crossed Andreev reflection can be a very effective tool in order
to identify the pairing symmetry of these superconductors.
\end{abstract}

\pacs{75.70.Cn,72.25Mk,74.90.+n} \keywords{Crossed Andreev
reflection, coexistence, pairing symmetry } \maketitle

{\em Introduction}: Recent experiments on certain materials like
$RuSr_{2}GdCu_{2}O_{8}$ and $UGe_{8}$ have indicated that the
previously thought to be antagonistic phenomenon of ferromagnetism
and superconductivity can and do coexist\cite{floquet}. This
suggests that the pairing symmetry of such ferromagnetic
superconductors may not be of the conventional BCS $s-wave$
singlet type. Indeed experiments in many rare earth and uranium
compounds provide quite distinct characteristics of triplet
superconductivity\cite{floquet}. This is of-course not conclusive
and more so in case of organic superconductors\cite{bolech}.
Currently there is much debate on the symmetry of the
super-conducting phase in organic compounds. There is of-course
lot of experimental evidence for the triplet pairing in
ruthenates. It is very unlikely for singlet superconductivity to
appear in the ferromagnetic state because the exchange interaction
forbids the formation of cooper pairs. Antiferromagnetic
correlations lead to singlet pairing (with zero spin), while
ferromagnetic correlations favor triplet pairing (with one unit of
spin). Another physical system where triplet pairing occurs is
superfluid Helium-3. Studies\cite{leggett} of $^{3}He$ suggest
that unconventional superconductivity will be highly anisotropic,
i.e., it will depend strongly on the energy and momentum of the
electrons. In $^{3}He$, the cooper pairs have an orbital angular
momentum and spin $1$. Spin $0$ objects have only one projection
($S_{z}=0$), spin $1$ objects have three projections ($S_{z}=-1,0$
or $1$). Superconductivity with spin $0$ cooper pairs is therefore
called singlet superconductivity and superconductivity with spin
$1$ cooper pairs is called triplet superconductivity. The singlet
state is asymmetric under exchange of spin labels while the
triplet state is symmetric. A magnetic field can destroy singlet
superconductivity is two ways. The first of these effects is known
as the orbital effect and is simply a manifestation of the lorentz
force. Since the electrons in the copper pair have opposite
momenta, the lorentz force acts in opposing directions and the
pair breaks up. The second phenomenon, known as the paramagnetic
effect, occurs when a strong magnetic field attempts to align the
spins of both the electrons along the field direction. Singlet
superconductivity is destroyed by fields greater than $\sim 1.8
T_{c}$, where $T_c$ is the critical temperature at which the
material loses its electrical resistance. Such fields, however do
not wreck triplet superconductivity because the spins of both
electrons may point in the same direction as the field. This means
that triplet superconductivity can only be destroyed by the
orbital effect.

The aim of this work is to provide a novel method of detection of
the pairing symmetry of ferromagnetic superconductors. Currently
the accepted and prevalent views on this subject are that spin
triplet pairing is the most likely symmetry. This assertion
of-course is qualified with the caveat that at weak magnetic
fields spin singlet pairing or non-uniform state (FFLO,
oscillating order parameter) can also be justified. This work does
not consider FFLO type states but it will be shown in the course
of this work that a clearer, simple and more intuitive distinction
between the singlet or triplet opposite spin pairing symmetry and
the triplet equal spin pairing symmetry can be easily brought
about by the phenomenon of crossed Andreev reflection(CAR). In
addition to the distinctive features observed in case of equal and
opposite spin pairing symmetries, the phases ($A_1$ and $A_2$)
seen for equal spin pairing can also be directly identified using
this phenomena\cite{bolech}. This problem has been dealt with in
literature by taking recourse to tunnelling spectroscopy, notably
in Ref.\cite{bolech}, where the effect of finite temperatures is
also taken into account. This problem has also been dealt with
through quantum pumping spectroscopy in Ref.\cite{tserko}.
Tunnelling spectroscopy in Ref.[2], is not as robust as the
phenomenon of CAR mentioned herein, since the distinction between
the $A_1$ and $A_2$ phases has not been brought out in Ref.[2].
Furthermore experiments of quantum pumping are notoriously
difficult and quantum pumping has not been unambiguously detected
as yet. The case of CAR is already clear cut,
theoretically(Ref.[4]) as well as experimentally(Ref.[6]). This
work uses the physics of CAR\cite{flatte}to identify correctly and
unambiguously the pairing symmetry of ferromagnetic
superconductors.

\begin{figure}[t]
\protect\centerline{\epsfxsize=2.0in\epsfbox{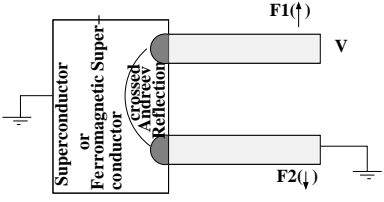}}
\caption{ Crossed Andreev reflection set-up. In the figure a bias
is applied to Ferromagnetic lead F1, while Superconductor (or,
Ferromagnetic Superconductor) and Ferromagnetic lead F2 are
grounded.}
\end{figure}

 To explain the phenomena of CAR\cite{deutscher_apl} we take recourse to Fig.~1, in
which two oppositely polarized ferromagnetic leads(half-metals)
are coupled to a singlet (opposite spin pairing) non-magnetic
superconductor. A voltage bias $eV_{} < \Delta$ (the
superconducting gap), is applied to the up-spin polarized
ferromagnetic lead F1, while the down-spin polarized ferromagnetic
lead F2 and the superconductor are grounded. In this case if the
distance between the leads is less than the superconducting
coherence length ($\xi$) then an up-spin electron incident in $F1$
with energy less than the superconducting gap will be completely
retro-reflected as a hole in $F2$ with spin down(Ref.[4]). This is
CAR. In the system as envisaged in Fig.~1 the barrier strengths at
the lead-superconductor interfaces: $z_{1}=z_{2}=0$. In this
set-up the formula for the current in leads F1 and F2 at zero
temperature are given as: $
I_{1}=I_{0}\int_{0}^{eV_{}}(1-B_{11})dE, \mbox {and }
I_{2}=I_{0}\int_{0}^{eV_{}}A_{12}dE$. In the preceding expressions
for the currents , $I_{0}=\frac{1}{2}N(0)ev_{F}{\cal A}$, with $
N(0)$ being the density of states, $v_{F}$ the Fermi velocity and
$\cal A$ the contact area between lead and superconductor
respectively. $B_{11}$ is the probability for normal reflection
and since applied bias is less than the superconducting gap and
$z_{1}=0$, $B_{11}=0$. $A_{12}$ is the probability for CAR which
in this case is $1$, since there are no down-spin states in F1,
the probability to be cross-reflected in F2 is unity. Thus,
$I_{1}=I_{2}=I=I_{0}V$. One can also derive the current in the
opposite case when a voltage bias $eV_{} < \Delta$ (the
superconducting gap) is applied to the down-spin polarized
ferromagnetic lead F2, while the up-spin polarized ferromagnetic
lead F1 and the superconductor are grounded. In this case too,
$z_{1}=z_{2}=0$. In this set-up the formula for the current in
leads F1 and F2 at zero temperature are given as: $
I_{1}=\int_{0}^{eV_{}}A_{21}dE, \mbox {and }
I_{2}=I_{0}\int_{0}^{eV_{}}(1-B_{22})dE, $
 $B_{22}$ of-course is the probability for normal reflection
and since applied bias is less than the superconducting gap and
$z_{2}=0$, $B_{22}=0$. $A_{21}$ is the probability for CAR which
in this case again is $1$ because of the same reason. Thus, again
$I_{1}=I_{2}=I=I_{0}V$.

 {\em Theory:} As we have seen in the last paragraph of the introduction
singlet s-wave superconductors show CAR. Now let us use the same
set-up as envisaged in Fig.~1, but now with the superconductor
replaced by a ferromagnetic superconductor. The hamiltonian of the
ferromagnetic superconductor can be easily diagonalized by a
Bogoliubov transformation and one can derive the
Bogoliubov-deGennes equation for the ferromagnetic superconductor
as written below. For the sake of brevity we only write below the
Bogoliubov-de Gennes equation as in Eq.~1. The interested reader
is referred to Ref.\cite{powell} for a full derivation of the
following equation starting from the hamiltonian of the
ferromagnetic superconductor.
\begin{widetext}
\begin{equation}
\left(\begin{array}{cccc}
\epsilon_{k}+h_{z} &0&-\Delta_{T,+1}&\Delta_{S}-\Delta_{T,0}\\
0&\epsilon_{k}-h_{z} &-\Delta_{S}-\Delta_{T,0}&-\Delta_{T,-1}\\
-\Delta_{T,+1}^{*}&-(\Delta_{S}^{*}+\Delta_{T,0}^{*})&-(\epsilon_{-k}+h_{z})&0\\
-(\Delta_{S}^{*}-\Delta_{T,0}^{*})&-\Delta_{T,-1}^{*}&0&-\epsilon_{-k}+h_{z}
\end{array} \right)\left(\begin{array}{c}
u_{\uparrow \sigma} \\
u_{\downarrow \sigma} \\
v_{\uparrow \sigma} \\ v_{\downarrow \sigma} \end{array}
\right)=E_{\sigma}\left(\begin{array}{c}
u_{\uparrow \sigma} \\
u_{\downarrow \sigma} \\
v_{\uparrow \sigma} \\ v_{\downarrow \sigma} \end{array} \right)
\end{equation}
\end{widetext}
 Herein we consider the magnetic field in the z-direction.The
meaning of the various symbols in Eq.~1 are given below:
$\Delta_{S}=\frac{1}{2}(\Delta_{\uparrow
\downarrow}-\Delta_{\downarrow \uparrow})$, singlet opposite spin
pairing; $\Delta_{T,0}=\frac{1}{2}(\Delta_{\uparrow
\downarrow}+\Delta_{\downarrow \uparrow})$, triplet opposite spin
pairing; $\Delta_{T,+1}=\Delta_{\uparrow \uparrow}$, triplet equal
spin pairing; $\Delta_{T,-1}=\Delta_{\downarrow \downarrow}$,
triplet equal spin pairing.

We analyze below four cases: (i) $\Delta_{S} \neq 0$, all others
are zero (singlet opposite spin pairing); (ii) $\Delta_{T,0} \neq
0$, all others are zero (triplet opposite spin pairing); (iii)
$\Delta_{T,+1} \neq 0, \Delta_{T,-1} \neq 0 $, all others are zero
[triplet equal spin pairing ($A2$ phase)]; and  (iv) Either
$\Delta_{T,+1} \neq 0$ and all others are zero  or $\Delta_{T,-1}
\neq 0$ and all others are zero, [triplet equal spin pairing ($A1$
phase)].

{\em Case 1: Singlet opposite spin pairing coexisting
with ferromagnetism}: In this case Eq.~1, can be separated into
two sets of equations:
\begin{eqnarray}
\left(\begin{array}{cc}
\epsilon_{k}+h_{z} &\Delta_{S}\\
-\Delta_{S}^{*}&-\epsilon_{-k}+h_{z}
\end{array} \right)\left(\begin{array}{c}
u_{\uparrow \uparrow} \\
v_{\downarrow \uparrow}  \end{array}
\right)=E_{\uparrow}\left(\begin{array}{c}
u_{\uparrow \uparrow} \\
v_{\downarrow \uparrow} \end{array} \right),\mbox{and}& &\\
\left(\begin{array}{cc}
\epsilon_{k}-h_{z} &-\Delta_{S}\\
-\Delta_{S}^{*}&-\epsilon_{-k}-h_{z}
\end{array} \right)\left(\begin{array}{c}
u_{\downarrow \downarrow} \\
v_{\uparrow \downarrow}  \end{array}
\right)=E_{\downarrow}\left(\begin{array}{c}
u_{\downarrow \downarrow} \\
v_{\uparrow \downarrow} \end{array} \right)& &
\end{eqnarray}
Solving Eqs.~2 and 3 for $E_\sigma$, we have:
$E_{\sigma}=\sqrt{\epsilon_{k}^{2}+|\Delta_{S}|^{2}}+\sigma h_z$
and the superconducting coherence factors $u_{\sigma \sigma},
v_{\sigma \sigma}$ are given by: $u_{\sigma
\sigma}^{2}=\frac{1}{2}(1+\frac{\epsilon_{k}}{E_{0}})$ and
$v_{\sigma
\overline\sigma}^{2}=\frac{1}{2}(1-\frac{\epsilon_{k}}{E_{0}})$.
These coherence factors have the same form as in the non-magnetic
superconductor. Hence, in the gap equation, exchange field enters
only in the fermi function and at zero temperature the gap
equation is independent of exchange field. Thus the results for
the non-magnetic case are valid in this case too at zero
temperature.

{\em Case $2$: Triplet opposite spin pairing
coexisting with ferromagnetism}:
 In this case Eq.~1 can be separated into two sets of equations:
\begin{eqnarray}
\left(\begin{array}{cc}
\epsilon_{k}+h_{z} &-\Delta_{T,0}\\
-\Delta_{T,0}^{*}&-\epsilon_{-k}+h_{z}
\end{array} \right)\left(\begin{array}{c}
u_{\uparrow \uparrow} \\
v_{\downarrow \uparrow}  \end{array}
\right)=E_{\uparrow}\left(\begin{array}{c}
u_{\uparrow \uparrow} \\
v_{\downarrow \uparrow} \end{array} \right), \mbox{and}& &\\
\left(\begin{array}{cc}
\epsilon_{k}-h_{z} &-\Delta_{T,0}\\
-\Delta_{T,0}^{*}&-\epsilon_{k}-h_{z}
\end{array} \right)\left(\begin{array}{c}
u_{\downarrow \downarrow} \\
v_{\uparrow \downarrow}  \end{array}
\right)=E_{\downarrow}\left(\begin{array}{c}
u_{\downarrow \downarrow} \\
v_{\uparrow \downarrow} \end{array} \right)& &
\end{eqnarray}
Solving Eqs.~4 and 5 for $E_\sigma$, we have:
$E_{\sigma}=\sqrt{\epsilon_{k}^{2}+|\Delta_{T,0}|^{2}}+\sigma h_z$
and the superconducting coherence factors $u_{\sigma \sigma},
v_{\sigma \overline\sigma}$ are given by: $u_{\sigma
\sigma}^{2}=\frac{1}{2}(1+\frac{\epsilon_{k}}{E_{0}})$ and
$v_{\sigma
\overline\sigma}^{2}=\frac{1}{2}(1-\frac{\epsilon_{k}}{E_{0}})$.
These coherence factors again have the same form as in the
non-magnetic superconductor. Thus in the gap equation, exchange
field enters only in the fermi function and at zero temperature
the gap equation is independent of exchange field. This is exactly
similar to the case of singlet opposite spin pairing coexisting
with ferromagnetism. Thus the results for the non-magnetic case
are valid in this case too at zero temperature. With this we go
over to the final case study on triplet equal spin pairing
symmetry.

{\em Case 3: Triplet equal spin pairing coexisting with
ferromagnetism:} In this case Eq.~1, can be separated into two
sets of equations:
\begin{eqnarray}
\left(\begin{array}{cc}
\epsilon_{k}+h_{z} &-\Delta_{T,+1}\\
-\Delta_{T,+1}^{*}&-\epsilon_{-k}-h_{z}
\end{array} \right)\left(\begin{array}{c}
u_{\uparrow \uparrow} \\
v_{\uparrow \uparrow}  \end{array}
\right)=E_{\uparrow}\left(\begin{array}{c}
u_{\uparrow \uparrow} \\
v_{\uparrow \uparrow} \end{array} \right)\mbox{and}& &\\
\left(\begin{array}{cc}
\epsilon_{k}-h_{z} &-\Delta_{T,-1}\\
-\Delta_{T,-1}^{*}&-\epsilon_{-k}+h_{z}
\end{array} \right)\left(\begin{array}{c}
u_{\downarrow \downarrow} \\
v_{\downarrow \downarrow}  \end{array}
\right)=E_{\downarrow}\left(\begin{array}{c}
u_{\downarrow \downarrow} \\
v_{\downarrow \downarrow} \end{array} \right) & &
\end{eqnarray}
 Solving Eqs. 6 and 7 for $E_\sigma$, we have:
$E_{\sigma}=\sqrt{(\epsilon_{k}+\sigma h_{z})^{2}+|\Delta_{\sigma
\sigma }|^{2}}$ and the superconducting coherence factors
$u_{\sigma \sigma}, v_{\sigma \sigma}$ are given by: $u_{\sigma
\sigma}^{2}=\frac{1}{2}(1+\frac{\epsilon_{k}+\sigma
h_{z}}{E_{\sigma}})$ and
$v_{\sigma\sigma}^{2}=\frac{1}{2}(1-\frac{\epsilon_{k}+\sigma
h_{z} }{E_{\sigma}})$. In this case of-course as is evident from
the coherence factors the exchange field does enter into the
factors. Further since spin-up and spin-down are completely
separated, there are two kinds of pairing state: (i) When only one
kind of pairing is non-zero, i.e., either $\Delta_{\uparrow
\uparrow} \neq 0$ or  $\Delta_{\downarrow \downarrow} \neq 0$,
which is identified as the $A1$ phase in the literature on spin
triplet materials and (ii) when both kinds of pairing are non-zero
which is identified as the $A2$ phase. One can as stated above
further categorize the $A1$ phase. When only $\Delta_{\uparrow
\uparrow} \neq 0$ and all others are zero we identify it as
$A1_\uparrow$ phase and when only $\Delta_{\downarrow \downarrow}
\neq 0$ and all others are zero we identify it as $A1_\downarrow$
phase. The above phases can be easily distinguished with the
method of CAR as is shown below.

 {\em  Case A: Triplet equal spin pairing $A1_\uparrow$ phase, with
$\Delta_{\uparrow
 \uparrow} \neq 0$ and $\Delta_{\downarrow \downarrow} = 0$ :} We
use the set-up as envisaged in Fig.~1. First we apply a bias $eV <
\Delta_{\uparrow \uparrow}$ to lead F1 and keep the
Ferromagnetic-Superconductor(FS) and lead F2 grounded. Since in
this case the pairing is only between two electrons with up-spin,
CAR into F2 is absent. There is only Andreev reflection (of a hole
with same up-spin as the incoming electron) into the same lead,
i.e., F1. Further since as before there are no interface barriers
at the junction between leads and superconductor, the normal
reflection at F1 is also absent. Thus, the currents in the two
leads are: $ I_{1}=I_{0}\int_{0}^{eV_{}}(1-B_{11}+A_{11})dE, \mbox
{and } I_{2}=I_{0}\int_{0}^{eV_{}}A_{12}dE $.
 Since applied bias is less than the superconducting gap and
$z_{1}=0, B_{11}=0$. $A_{11}$ the probability for normal
reflection is $1$. Further since there is complete absence of CAR
$A_{12} = 0$. Thus, $I_{1}=2I=2I_{0}V$ and $I_{2}=0$. One can also
derive the current in the opposite case when a voltage bias $eV_{}
< \Delta_{\uparrow \uparrow}$ (the $A1_\uparrow$ phase
superconducting gap) is applied to lead F2, while  lead F1 and FS
are grounded. In this case as before, $z_{1}=z_{2}=0$. In this
set-up, the formula for the current in leads F1 and F2 at zero
temperature are given as: $
I_{1}=I_{0}\int_{0}^{eV_{}}A_{21}dE,\mbox {and }
I_{2}=I_{0}\int_{0}^{eV_{}}(1-B_{22}+A_{22})dE $.
 Since applied bias is less than the superconducting gap and
$z_{2}=0, B_{22}=0$. $A_{22}$ is the probability for normal
Andreev reflection which in this case is zero, since there are no
up-spin states in F2 to get reflected into. Further the
probability for CAR $A_{21}$ is also nil, since in the
$A1_\uparrow$ phase envisaged $\Delta_{\downarrow \downarrow}=0$.
This is because of the fact that F2 is completely down-spin
polarized. Thus, $I_{1}=0 \mbox {
 and } I_{2}=I=I_{0}V$.

{\em Case B: Triplet equal spin pairing $A1_\downarrow$ phase,
with $\Delta_{\uparrow \uparrow} = 0$ and $\Delta_{\downarrow
\downarrow} \neq 0$:} We use the set-up as envisaged in Fig.~1.
First we apply a bias $eV < \Delta_{\downarrow \downarrow}$ to
lead F1 and keep the FS and lead F2 grounded. Since in this case
the pairing is only between two electron with down-spin, CAR into
F2 is absent, for the same reasons Andreev reflection (of a hole
with same up-spin as the incoming electron) into the same lead,
i.e., F1 is also absent. Further since as before there are no
interface barriers at the junction between leads and
superconductor, the normal reflection at F1 is also absent. Thus,
the currents in the two leads are: $
I_{1}=I_{0}\int_{0}^{eV_{}}(1-B_{11}+A_{11})dE,\mbox {and }
I_{2}=I_{0}\int_{0}^{eV_{}}A_{12}dE $.
 Since applied bias is less than the superconducting gap and
$z_{1}=0$, $B_{11}=0$. $A_{11}$ is zero, since pairing only
involves down-spin electrons. Further since there is complete
absence of CAR, $A_{12} = 0$. Thus, $I_{1}=I=I_{0}V$ and
$I_{2}=0$. One can also derive the current in the opposite case
when a voltage bias $eV_{} < \Delta_{\downarrow \downarrow}$ (the
$A1_\downarrow$ phase superconducting gap) is applied to lead F2,
while lead F1 and the FS are grounded. In this case as before,
$z_{1}=z_{2}=0$. For this set-up the current in leads F1 and F2 at
zero temperature are given as: $
I_{1}=I_{0}\int_{0}^{eV_{}}A_{21}dE,\mbox {and }
I_{2}=I_{0}\int_{0}^{eV_{}}(1-B_{22}+A_{22})dE $.
 Since applied bias is less than the superconducting gap and
$z_{2}=0$, $B_{22}=0$. $A_{22}$ is $1$. Further the probability
for CAR $A_{21}$ is nil, since the $A1_\downarrow$ phase
envisages, $\Delta_{\uparrow \uparrow}=0$. This is of-course due
to the fact that F2 is completely down-spin polarized, and there
are no up-spin states in F2. Thus, $I_{1}=0 \mbox{ and }
I_{2}=2I=2I_{0}V$. At the end we analyze the case wherein both
$\Delta_{\uparrow \uparrow}\neq 0$ and $\Delta_{\downarrow
\downarrow} \neq 0$, i.e., the $A2$ phase.

{\em Case C: Triplet equal spin pairing $A2$ phase, with
$\Delta_{\uparrow \uparrow} \neq 0$ and $\Delta_{\downarrow
\downarrow} \neq 0$:} We use the set-up as envisaged in Fig.~1.
First we apply a bias $eV < \Delta_{\downarrow \downarrow} ( \mbox
{and }\Delta_{\uparrow \uparrow})$ to lead F1 and keep FS and lead
F2 grounded. It should be noted that the applied bias should be
less than both $\Delta_{\uparrow \uparrow}$ and
$\Delta_{\downarrow \downarrow}$. Since in this case the pairing
is both between two electrons with down-spin as well as between
two electrons with up-spin, Andreev reflection (of a hole with
same up-spin as the incoming electron) into the same lead, i.e.,
F1 is present. Further since as before there are no interface
barriers at the junction between leads and superconductor, the
normal reflection at F1 is absent. As far as CAR $A_{12}$ into F2
is concerned, since F2 is completely down spin polarized and in
this case incident electron with up-spin is incident in F1,
$A_{12}=0$. Thus, the currents in the two leads are: $
I_{1}=I_{0}\int_{0}^{eV_{}}(1-B_{11}+A_{11})dE,\mbox {and }
I_{2}=I_{0}\int_{0}^{eV_{}}A_{12}dE $.
 Since applied bias is less than the super-conducting gap and
$z_{1}=0$, $B_{11}=0$. $A_{11}$ in this case is $1$, while as
discussed above $A_{12}=0$. Thus, $I_{1}=2I=2I_{0}V$ and
$I_{2}=0$. One can also derive the current in the opposite case
when a voltage bias $eV_{} < \Delta_{\downarrow \downarrow}( \mbox
{and }\Delta_{\uparrow \uparrow})$ (the super-conducting gap) is
applied to lead F2, while F1 and FS are grounded. In this case
too, $z_{1}=z_{2}=0$. In this set-up the formula for the current
in leads F1 and F2 at zero temperature are thus given as: $
I_{1}=I_{0}\int_{0}^{eV_{}}A_{21}dE,\mbox {and }
I_{2}=I_{0}\int_{0}^{eV_{}}(1-B_{22}+A_{22})dE $.
 Since applied bias is less than the super-conducting gap and
$z_{2}=0$, $B_{22}=0$. $A_{22}$ in this case is $1$. Further the
probability for CAR $A_{21}$ is nil, since F1 is purely up-spin
polarized and there are no down-spin states. Hence, $I_{1}=0,
\mbox {and } I_{2}=2I=2I_{0}V$. Thus, the important message we get
from the above analysis of singlet opposite spin pairing case and
of the triplet opposite spin pairing case is not whether the
pairing symmetry is triplet or singlet but whether it is opposite
spin pairing or equal spin pairing which really matters.

{\em Experimental Realization:}  The phenomenon of CAR has been
demonstrated in two recent experiments\cite{beckmann}. To adapt it
so that it can reveal the pairing symmetry of FS one has to
replace the normal superconductor used in the aforementioned
experiments with the FS. After this the two oppositely polarized
Ferromagnetic leads are half metals. A proper choice for the
purely up-spin polarized Ferromagnetic lead F1 could be Chromium
Oxide $CrO_2$, while for the purely down-spin polarized
Ferromagnetic lead F2 could be the Strontium alloy
$Sr_{2}FeMoO_6$.

\begin{table}[h]
\squeezetable
\begin{center}
\caption{\small{Distinguishing opposite spin-pairing and
equal-spin pairing in Ferromagnetic Superconductors}}\vspace{0mm}
\begin{tabular}{|c|c|c|c|c|}
\hline
  & {\em Opposite }&\multicolumn{3}{c|}{\em Equal Spin-Pairing (Triplet)}\\ \cline{3-5}
 &{\em Spin-Pairing}& \multicolumn{2}{c|}{A$1$ Phase} & A$2$ Phase\\ \cline{3-4}
  &  (Singlet/Triplet)                 &$A1_{\uparrow}(\Delta_{\uparrow \uparrow}\neq 0)$&$A1_{\downarrow}(\Delta_{\downarrow
  \downarrow}\neq 0)$& \\ \hline
 Bias& $I_{1}=I$&$I_{1}=2I$&$I_{1}=I$&$I_{1}=2I$ \\
applied &$I_{2}=I$ &$I_{2}=0$ & $I_{2}=0$&$I_{2}=0$\\
to F1&$dG_{T}=0$ &$dG_{T}=I_0$ & $dG_{T}=I_{0}/2$&$dG_{T}=I_0$\\
&$dG_{G}=2I_0$ &$dG_{G}=2I_0$&$dG_{G}=I_{0}$&$dG_{T}=2I_0$\\\hline
Bias & $I_{1}=I$ &$I_{1}=0$&$I_{1}=0$& $I_{1}=0$ \\
applied &$I_{2}=I$&$I_{2}=I$ &$I_{2}=2I$ &$I_{2}=2I$ \\
to F2&$dG_{T}=0$&$dG_{T}=-I_{0}/2$&$dG_{T}=-I_{0}$&$dG_{T}=-I_{0}$\\
&$dG_{G}=2I_0$&$dG_{G}=I_{0}$&$dG_{G}=2I_{0}$&$dG_{G}=2I_{0}$\\\hline
\end{tabular}
\end{center}
\end{table}

{\em Conclusions:} Finally we juxtapose all the results obtained
in this work in Table 1. We see that the currents for all the
cases bear the characteristic of the pairing symmetry considered.
Especially one can clearly mark out the characteristic differences
between $A1$ and $A2$ phases. Furthermore one can also make out
which pairing (up/down) is non-zero in the $A1$ phase. This makes
the phenomena of CAR ideal for the detection of the pairing
symmetry of Ferromagnetic Superconductors. We also calculate the
differential conductance $dG=dI/dV$ for the different symmetry
classes. The differential conductance when bias is applied to F1
is just current in F2 divided by the voltage, and when bias is
applied to F2 is just current in F1 divided by the voltage which
can be inferrred from table 1 by replacing I in the $I_{1}/I_{2}$
currents (depending upon to which lead bias is applied) by
$I_{0}$. Two interesting quantities labelled in Ref.[3] as
$I_{T}=I_{1}-I_{2}$ and $I_{G}=I_{1}+I_{2}$, the differential
conductance from these quantities are $dG_{T}=dI_{T}/dV$ and
$dG_{G}=dI_{G}/dV$, these quantities can also reveal the
differences between the different symmetry classes as again
pointed out in table 1.

 A perspective on the application is called for. First, if ferromagnets
are not half-metals, then the situation changes depending on the
polarization strength of the ferromagnets, see
Ref.\cite{yamashita}. The effects of impurities have been dealt
with in Ref.\cite{melin}, where they showed that novel features
like reflectionless tunnelling appear. This suggests CAR will be
contaminated by elastic co-tunneling when ferromagnets are not
half-metals. For half-metals impurities will not change anything
qualitatively, i.e., when currents are zero they remain zero.
Lastly temperature also wont change anything qualitatively, since
temperature only appears in the fermi functions. To conclude, one
can say for half-metals, impurities or finite temperature will not
affect the currents qualitatively.

\end{document}